\documentclass[12pt]{article}

\global\arraycolsep=1pt
\usepackage{amssymb}
\usepackage{amsmath}
\usepackage{amsthm}
\usepackage{amscd}
\usepackage {pstricks}
\allowdisplaybreaks
\newcommand{\beq}{\begin{equation}}
\newcommand{\eeq}{\end{equation}}
\newcommand{\beqa}{\begin{eqnarray}}
\newcommand{\eeqa}{\end{eqnarray}}
\newcommand{\CR}{\nonumber \\}

\newcommand{\m}{\mu}

\newcommand{\C}{{\mathbb C}}
\newcommand{\R}{{\mathbb R}}
\newcommand{\Z}{{\mathbb Z}}

\newcommand{\diag}{\mathrm{diag~}}
\newcommand{\trace}{\mathrm{Tr~}}

\newcommand{\unitbox}
{\setlength{\unitlength}{0.5pt}
\begin{picture}(10,10)
\put(0,10){\line(1,0){10}}
\put(0,0){\line(1,0){10}}
\put(0,0){\line(0,1){10}}
\put(10,0){\line(0,1){10}}
\end{picture}}

\renewcommand{\thefootnote}{\fnsymbol{footnote}}

\theoremstyle{definition}
\newtheorem{thm}{Theorem}

\begin{document}

\begin{titlepage}
\begin{flushright}
{\tt hep-th/0702125} 
\end{flushright}
\vspace{0.5cm}
\begin{center}
{\Large \bf Instanton calculus and chiral one-point functions\\
in supersymmetric gauge theories}
\vskip1.5cm
{\large Shigeyuki Fujii, Hiroaki Kanno,  \\
Sanefumi Moriyama and Soichi Okada}
\vskip 2em
{\it 
Graduate School of Mathematics \\
Nagoya University, Nagoya, 464-8602, Japan}
\end{center}
\vskip2cm

\begin{abstract}
We compute topological one-point functions of the chiral operator $\trace \varphi^k$
in the maximally confining phase of $U(N)$ supersymmetric gauge theory. 
These one-point functions are polynomials in the equivariant parameter $\hbar$ 
and the parameter of instanton expansion $q=\Lambda^{2N}$ and are of particular
interest from gauge/string theory correspondence, since they are related to
the Gromov-Witten theory of ${\bf P}^1$. Based on a combinatorial identity that
gives summation formula over Young diagrams of relevant functions,
we find a relation among chiral one-point functions, which recursively 
determines the $\hbar$ expansion of the generating function of one-point functions.
Using a result from the operator formalism of the Gromov-Witten theory,
we also present a vacuum expectation value of the loop operator 
$\trace e^{it\varphi}$.

\end{abstract}
\end{titlepage}


\renewcommand{\thefootnote}{\arabic{footnote}}
\setcounter{footnote}{0}


\section{Introduction}
\setcounter{equation}{0}

Recently there is a substantial progress in the instanton calculus of
four dimensional gauge theories \cite{Nek, FP, BFMT, LMN, NO, NY1, FFMP, FMPT}. 
In particular, Nekrasov proposed 
a partition function $Z_{\mathrm Nek}(\epsilon_i, a_\ell, \Lambda)$ 
that encodes the information of the instanton counting in four dimensional
gauge theory. In \cite{Nek} integrations over the instanton
moduli space are evaluated by equivariant localization principle,
where the equivariant parameters $(\epsilon_1, \epsilon_2)$ of the toric
action on ${\mathbb C}^2 \simeq {\mathbb R}^4$ can be identified
as those of the spacetime non-commutativity, 
or physically the graviphoton background.
The fixed points of the toric action are labeled by the partitions
or in other words the Young diagrams. Consequently the non-perturbative partition
function and correlation functions are expressed as
summations of the functions on the set of Young diagrams.
We can show that a five dimensional lift
 (or \lq\lq trigonometric\rq\rq\  lift) of Nekrasov's 
partition function $Z_{\mathrm {Nek}}^{\mathrm {5D}}$ is nothing but the partition 
function of topological string (the generating function of Gromov-Witten
invariants) $Z_{\mathrm {top~str}}^{(K_s)}$ on a local toric Calabi-Yau 3-fold
$K_S$, where $S$ is an appropriate toric surface \cite{IK1, IK2, EK1, EK2, Zhou}.
The correspondence of Nekrasov's partition function and the generating
function of the Gromov-Witten invariants of local Calabi-Yau manifold,
$Z_{\mathrm {Nek}}^{\mathrm {5D}} \equiv Z_{\mathrm {top~str}}^{(K_s)}$, 
is one example of gauge/string correspondence in topological
theory \cite{GV}, which is expected from 
the idea of geometric engineering \cite{KKV}.

In this paper, we explore another example of 
gauge/string correspondence which involves the topological one-point functions. 
In $U(N)$ supersymmetric gauge theory in four dimensions, 
there are chiral observables $\trace \varphi^k$, where $\varphi$ is 
the (Higgs) scalar field in the adjoint representation.
We will present a result on the computation of the vacuum
expectation value of $\trace \varphi^k$ in the $U(1)$ gauge theory or
in the maximally confining phase of $U(N)$ theory,
where the effective low energy symmetry is reduced to $U(1) \subset U(N)$.
In \cite{LMN} A. Losev, A. Marshakov and N. Nekrasov
claimed that there is a gauge/string correspondence
${\trace} \varphi^{2j} \Longleftrightarrow \tau_p(\omega)$,
where ${\trace} \varphi^{2j}$ are generators of the chiral ring and 
$\tau_p(\omega)$ is the $p$-th gravitational descendant of 
the K\"ahler class $\omega$ of ${\bf P}^1$. Thus it is expected that
the correlation functions of chiral ring elements are related to
the Gromov-Witten invariants of ${\bf P}^1$ developed 
by Okounkov and Pandharipande \cite{OP1,OP2}.

The one-point functions $\langle{\trace} \varphi^{2j}\rangle$
are polynomials in the parameter of instanton expansion $q:=\Lambda^{2N}$ 
and the equivariant parameter of the toric action $\hbar = \epsilon_1
= -\epsilon_2$. In the gauge/string correspondence
these parameters play complementary roles. For example,
Nekrasov's partition function allows the following two kinds of expansion;
\beq
Z_{\mathrm Nek}(\hbar, a_\ell,\Lambda) 
= \sum_{k=0}^\infty \Lambda^{2N \cdot k}~Z_k(\hbar, a_\ell) 
= \exp \left( -  \sum_{r=0}^\infty \hbar^{2r-2}  F_r(a_\ell, \Lambda) \right)~. \label{expansion}
\eeq
In gauge theory we primarily want to sum up the instanton
expansion in $\Lambda$, which is achieved for example in Seiberg-Witten theory.
On the other hand the expansion in $\hbar$ is identified with the genus expansion
in the corresponding (topological) string theory. The genus zero part
$F_0(a_\ell, \Lambda)$ gives the prepotential of Seiberg-Witten theory and 
higher order terms are expected to represent gravitational corrections 
\cite{Nek, KMT, DST, EK1, BFFL}.

In this paper we first take the viewpoint of gauge theory and 
the one-point functions $\langle{\trace} \varphi^{2j}\rangle$ are defined
not for each fixed instanton number but by summing up all the instanton numbers. 
We emphasize this point, 
since the partition function $Z_{U(1)}$ that appears in the definition of one-point functions 
contains the contributions from all the instanton numbers. 
We show that the chiral one-point functions
$\langle{\trace} \varphi^{2j}\rangle$ satisfy the relation
\begin{align}
\sum_{j=1}^r c_j^r \hbar^{2(r-j)}\langle\trace\varphi^{2j}\rangle
=\frac{(2r)!}{(r!)^2}q^r~,
\label{main}
\end{align}
which is one of the main results in the paper.
The coefficients $c_j^r$ are defined by $\prod_{j=0}^{r-1} (x^2 -j^2) = \sum_{j=1}^r c_j^r x^{2j}$ or
a specialization of the elementary symmetric functions $e_n(x)$; $c_j^r = (-1)^{r-j} e_{r-j} (1^2, 2^2, \cdots, (r-1)^2)$.
From the above linear relations \eqref{main} among one-point functions, we can compute
the expansion in $\hbar^2$ of the generating function $T(z)$ of 
one-point functions $\langle{\trace} \varphi^{2j}\rangle$, order by order.
Technically, our proof of the relation \eqref{main} is based on combinatorial
identities, which we obtain by considering the power sums of Jucys-Murphy 
elements in the class algebras of symmetric groups.

Complementary to the above computation is 
the computation of the Gromov-Witten invariants in \cite{OP1,OP2} by operator method, 
which gives all genus results for each fixed instanton number on the gauge theory side. 
In this sense the operator formalism naturally provides the generating function of the
identities \eqref{main} for each instanton sector. Summing up all the instanton numbers,
we can calculate the vacuum expectation value of the loop operator without difficulty.
Our final result is
\begin{align}
\langle\trace e^{it\varphi}\rangle
=I_0\bigl(2\sqrt{q}\, {\mathrm{sh}}(it\hbar)/\hbar\bigr)~,
\end{align}
with $ {\mathrm{sh}} (z)=e^{\frac{z}{2}}-e^{-\frac{z}{2}}$ and $I_n(x)$ being the
modified Bessel functions. It is remarkable that
the modified Bessel functions appear frequently in the
computation of the correlation functions of the loop operator
\cite{MSS,ESZ,DG,BH}. In our case $\langle\trace e^{it\varphi}\rangle
= I_0(2i\sqrt{q}~t)$ when $\hbar \to 0$ and  the effect of the equivariant deformation by $\hbar$ 
is taken care of simply by renormalizing the parameter $t$
as $it \to  {\mathrm{sh}} (it\hbar)/\hbar$.

The paper is organized as follows. In section 2
we review the basic tools in instanton calculus;
the ADHM construction of the instanton moduli space and localization 
formula concerning the toric action on the moduli space. 
In section 3 we consider the one-point function 
$\langle{\trace} \varphi^{2j}\rangle$ and derive \eqref{main}. 
The genus expansion of the generating function $T(z)$ is worked out in section 4. 
Computation in the operator formalism 
and comparison with the Gromov-Witten theory are made in section 5. 
The generating function of the relation \eqref{main} is naturally 
related to the loop operator and 
we calculate its vacuum expectation value in section 6. 
Finally we prove a crucial combinatorial formula in Appendix. 

It has been argued that the generating function of the Gromov-Witten invariants of
${\bf P}^1$ is a tau-function of Toda lattice hierarchy \cite{OP1,OP2}. 
In this paper we have obtained the $\hbar$ expansion of the generating function $T(z)$ 
of chiral one-point functions.  It is interesting to clarify a relation of this genus expansion
to integrable hierarchy and matrix models. For a recent paper in this direction, see \cite{MN}.


\section{ADHM construction and localization formula}
\setcounter{equation}{0}

For describing the moduli space of instantons, there is  
a strong tool called ADHM construction.
The instanton moduli space ${\mathcal M}_{N,k}$
of $U(N)$ gauge theory on ${\C}^2$ with instanton number $k$ is constructed
by introducing matrices\footnote{$M_{\C}(m,n)$ denotes the set of
$m \times n$ complex matrices.}
$B_1 , B_2 \in M_{\C}(k,k),$ $J \in M_{\C}(N,k)$ 
and $I \in M_{\C}(k,N)$ on $\C$.
Combining these matrices and coordinates $(z_1 , z_2 )$ of ${\C}^2$ , 
we define an $(N+2k) \times 2k$ matrix
\beqa
	\Delta :=\left(
		\begin{array}{ccc}
			J & &I^{\dagger } \\
			B_1 -z_1&~~&-B_2^{\dagger }+\overline{z_2} \\
			B_2 -z_2&~~&B_1^{\dagger }-\overline{z_1}
		\end{array}
	\right).
\eeqa
We construct an $(N+2k) \times N$ matrix $U$, whose column vectors consist of
a basis of the kernel of $\Delta$, i.e. a matrix $U$ that satisfies 
$\Delta^{\dagger } U=0$. 
A $U(N)$ connection $A$ is defined by $A:=U^{\dagger } (z) d_{{\C}^2} U(z)$. 
Then from the self-duality of $A$ and the normalization condition on $U$,
we obtain ADHM equations,
\beqa
	\left\{
		\begin{array}{lll}
			&\mu_{\C} &:=[B_1 , B_2 ]+IJ = 0\,,\\
			&\mu_{\R} &:=[B_1 , B_1^{\dagger }]+[B_2 , B_2^{\dagger }]+II^{\dagger }-J^{\dagger} J = 0\,.
		\end{array}
	\right. \label{ADHM}
\eeqa
Elements in $\mu_{\C}^{-1} (0) \cap \mu_{\R}^{-1} (0)$ give 
$k$-instantons of the $U(N)$ gauge theory on $\C^2$.

Since ADHM equations \eqref{ADHM} are invariant under the action
\beq
(B_1 , B_2 , J , I) \mapsto (T_{\phi}^{-1} B_1 T_{\phi} \, 
, \, T_{\phi}^{-1} B_2 T_{\phi} \, , \, JT_{\phi} \, , \,T_{\phi}^{-1} I)~,
\eeq
of $T_{\phi }=\exp (i\phi ) \in U(k)$, we may consider the quotient 
\beq
{\mathcal M}_{N,k}^0 := \mu_{\C}^{-1} (0)\cap \mu_{\R}^{-1} (0)/U(k)~,
\eeq
which is isomorphic to the moduli space of $k$-instantons of the $U(N)$ gauge theory.
However, in general ${\mathcal M}_{N,k}^0$ is singular 
and we consider a smooth manifold 
$${\mathcal M}_{N,k}^{\zeta} :=\mu_{\C}^{-1} (0)\cap \mu_{\R}^{-1} (\zeta )/U(k)~,$$
instead, as a resolution of ${\mathcal M}_{N,k}^0$. 
We note that there are several viewpoints 
on the manifold ${\mathcal M}_{N,k}^{\zeta}$. Each viewpoint has its own
advantages. Firstly, it can be regarded as
the moduli space of instantons on the non-commutative $\C^2$, where
$\zeta$ corresponds to the non-commutative parameter,
$[z_1 ,\overline{z_1} ]=-\zeta /2 \, , \, [z_2 ,\overline{z_2} ]=-\zeta /2.$
One can also regard ${\mathcal M}_{N,k}^{\zeta}$ as 
the moduli space of framed torsion free sheaves $(E,\Phi)$ on ${\bf P}^2$,
where $E$ is a torsion free sheaf of rank $N$ with
$\langle c_2 (E),[{\bf P}^2] \rangle=k$ and
locally free at a neighborhood of $l_{\infty}$ (the line at infinity) and 
$\Phi$ is an isomorphism $\Phi:E|_{l_{\infty}} 
\longrightarrow {\mathcal O}_{l_{\infty}}^{\oplus N}$, called framing operator.  
Finally, if we set the rank $N$ of gauge group to $1$, ${\mathcal M}_{1,k}^{0}$ 
and ${\mathcal M}_{1,k}^{\zeta}$ are isomorphic to
the symmetric product $S^k (\C^2)$ of $\C^2$ and the
Hilbert schemes of points $(\C^2)^{[k]}$ on $\C^2$, respectively. 

In four dimensional gauge theory with the gauge group $U(N)$,
we can consider the two kinds of toric action;
\begin{itemize}
\item
$\xi_{\C^2}$ action on $(z_1, z_2) \in {\C}^2$ defined by
$(z_1, z_2) \to (e^{i \epsilon_1} z_1, e^{i \epsilon_2} z_2)$.
Physically this introduces a constant (electro-magnetic) flux 
on ${\C}^2$ and makes it {\it non-commutative}.
This is called \lq\lq$\Omega$ background\rq\rq\ 
of Nekrasov. 
In the following we often put $\hbar = \epsilon_1 = -\epsilon_2$
(this is the self-duality or \lq\lq Calabi-Yau\rq\rq\ condition).
\item
The action of the maximal torus $(e^{i a_1}, \ldots, e^{i a_{N}}) \in U(1)^{N}$  on $U(N)$. 
In ${\mathcal N}=2$ supersymmetric gauge theory, the corresponding equivariant parameters 
$a_\ell$ are identified as vacuum expectation values of Higgs scalar in the vector multiplet.
\end{itemize}

These toric actions induce the following action of 
$T=U(1)^2 \times U(1)^N$ on the instanton moduli space ${\mathcal M}_{N,k}^{\zeta}$,
which allows us to employ a powerful tool of localization formula;
\beq
\xi:(B_1 , B_2 ,I,J) \mapsto (T_{\epsilon_1} B_1 ,T_{\epsilon_2} B_2 , IT_a^{-1} , T_{\epsilon_1} T_{\epsilon_2} T_a J)~,
\eeq
where $T_{\epsilon_k} := e^{i \epsilon_k} \in U(1)$ and $T_a := \diag (e^{i a_1} ,\ldots ,e^{i a_N}) \in U(1)^N$.
Let $\xi$ denote the vector field associated with the toric action. 
The equivariant differential operator $d_{\xi} :=d_{\C^2} + d_{{\mathcal M}} - \iota_{\xi}$ on 
$\Omega^{\bullet} ( \C^2  \times {\mathcal M}_{N,k}^{\zeta} )\otimes \C [\mathfrak{g}]$ satisfies
$d_{\xi}^2 = - {\mathcal L}_{\xi}$, 
where ${\mathcal L}_{\xi}$ is the Lie derivative associated to the action $\xi$.
We define ${\mathcal A}:=U^{\dagger} d_{\xi} U$ 
and ${\mathcal F}:=d_{\xi} (U^{\dagger} d_{\xi} U)$.
Mathematically ${\mathcal A}$ defines a connection on a rank $N$ vector bundle ${\mathcal E}$ 
on $\C^2 \times {\mathcal M}_{N,k}^{\zeta}$, called universal bundle and ${\mathcal F}$ is the curvature of
${\mathcal A}$. This identification was first provided in topological gauge theory 
\cite{BS, Kan}.
Since $\iota_{\xi} U =0$, we obtain the following decomposition concerning the direct product
$\C^2 \times {\mathcal M}_{N,k}^{\zeta}$;
\beqa
	\begin{array}{ll}
		{\mathcal A} &= U^{\dagger} d_{\C^2} U + U^{\dagger} d_{{\mathcal M}} U =: A+C~, \\
		{\mathcal F} &= d_{\C^2} (U^{\dagger} d_{\C^2} U)+ d_{\C^2} (U^{\dagger} d_{{\mathcal M}} U) +
			d_{{\mathcal M}} (U^{\dagger} d_{\C^2} U) +(d_{{\mathcal M}} U^{\dagger} )(d_{{\mathcal M}} U) 
			- U^{\dagger} {\mathcal L}_{\xi} U \\
		&= F_{\mu \nu} dx^{\mu} dx^{\nu} + \{ \lambda_m dz^m + \psi_{\overline{m}} d\overline{z}^{\overline{m}} \}
			+\{ (d_{{\mathcal M}} U^{\dagger} )(d_{{\mathcal M}} U) 
			- U^{\dagger} {\mathcal L}_{\xi} U \} \\
		&=: F + \Psi + \varphi~.
	\end{array}
\eeqa
In ${\mathcal N}=1$ supersymmetric Yang-Mills theory,
the components $A$, $F$, $\lambda_m dz^m$, 
$\psi_{\overline{m}} d{\overline{z}}^{\overline{m}}$ 
and $\varphi$ are identified with the gauge connection, 
the field strength (curvature), gaugino, chiral matter field 
and scalar field, respectively \cite{FMPT}.
We can see easily that it is only the scalar $\varphi$ that depends on $\xi$.

The topological partition function and correlation functions are defined by 
the equivariant integration on  ${\mathcal M}_{N,k}^{\zeta}$ and they are Laurent
series in the equivariant parameters $\epsilon_i$ and $a_\ell$. 
Namely, the correlator $\langle {\mathcal O} \rangle$ 
of an operator ${\mathcal O}$ is defined by
\beq
\langle {\mathcal O} \rangle
:= \frac{1}{V{\mathcal Z}} \int_{{\mathcal M}} \left\{ \int_{\C^2} {\mathcal O} \right\} 
\exp (-{\mathcal S}_{{\mathcal N}=1})~, \label{correldef}
\eeq
where the action for ${\mathcal N}=1$ supersymmetric gauge theory 
${\mathcal S}_{{\mathcal N}=1}$ is defined by
that of ${\mathcal N}=2$ theory ${\mathcal S}_{{\mathcal N}=2}$ perturbed 
by a superpotential $W(\Phi )$.
The correlator $\langle {\mathcal O} \rangle$ is normalized by
the volume $V$ of the non-commutative $\C^2$ and the partition function
${\mathcal Z}:=\int_{{\mathcal M}} \exp (-{\mathcal S}_{{\mathcal N}=1})$. 
The integral is over ${\mathcal M} :=\sqcup_k {\mathcal M}_{N,k}^{\zeta}$, that is, 
when we compute the correlation function we take a sum over
the instanton number $k$. 
Since they are computed by the instanton calculus that employs
the equivariant cohomology and the localization formula for the equivariant integral, 
let us first review the localization formula briefly.

Let $M$ be a smooth manifold of dimension $2l$ acted by a compact Lie group $G$.
The vector field associated to the $G$-action is denoted by $\xi$.
For the $G$-fixed point set $\Omega^p (M)^{\xi} := \{ \sigma \in \Omega^p (M) | {\mathcal L}_{\xi} \sigma =0 \}$ 
of $p$-forms on $M$, an element of $\Omega^{\bullet } (M)^{\xi} \otimes
\C [\mathfrak{g}]$ 
is called equivariant differential form associated to the vector field $\xi$, 
where $\mathfrak{g}$ is the Lie algebra of $G$.
Then we can define the cohomology $H^p_{\xi} (M)$, which 
is called equivariant de Rham cohomology, 
for equivariant differential forms using the differential operator $d_{\xi} := d-\iota_{\xi}$.
An equivariant differential form $\mu \in \Omega^p (M)^{\xi} \otimes  \C [\mathfrak{g}]$ 
is called equivariantly exact (resp. closed), if $\mu$ is written as $d_{\xi} \nu$ using 
an equivariant form $\nu \in \Omega^{p-1} (M)^{\xi} \otimes  
\C [\mathfrak{g}]$ (resp. $d_{\xi} \mu =0$).
The integral of equivariant forms on $M$
\beq
	\int_{M} : \Omega^{\bullet } (M)^{\xi} \otimes  \C [\mathfrak{g}] \longrightarrow \C [\mathfrak{g}]
\label{integral}
\eeq
defines a homomorphism and is called equivariant integral.
For calculating the equivariant integral, we can use a very powerful formula of localization \cite{BGV, GS};

\begin{thm}[Localization formula]
	If all fixed points of the $G$-action on $M$ are isolated, the integral of an equivariantly closed form
	$\mu$ is given by
	\beq
		\int_M \mu = (-2\pi )^l \sum_{s \in M^G} \frac{\mu_0 (s)}{\det^{\frac{1}{2}} {\mathcal L}_{\xi} (s)}~,
	\eeq
	where ${\mathcal L}_{\xi}$ is the homomorphism 
	${\mathcal L}_{\xi_i}^j := \partial \xi_i /\partial x^j :T_s M \longrightarrow T_s M$,
	$\mu_0$ is the zero-form part of $\mu$ and
	$M^G$ is the $G$-fixed points set on $M$.
\end{thm}
	
When the group $G$ is $U(1)^r$, $\det^{\frac{1}{2}} {\mathcal L}_{\xi} (s) = \prod_i (k_i (s) \cdot  \epsilon )$,
where $( k_1 (s), \ldots ,k_l (s) ) \in (\Z^r )^l$ are the weights of the representation of $U(1)^r$ 
at $s\in M^G$ and $\epsilon $ is the generator of $\mathfrak{g}$. Instanton part ${\mathcal Z}^{\mathrm{inst}}:= 
\sum_{k=0}^{\infty} q^k \int_{{\mathcal M}_{k,N}^{\zeta}} 1$ 
of Nekrasov's partition function ${Z}_{\mathrm{Nek}}$ for ${\mathcal N}=2$ super Yang-Mills theory
can be obtained by the equivariant integration of $ ``1"$
on the moduli space ${\mathcal M}$ of instantons on ${\C}^2$.
By the work of Nakajima \cite{Nak}, the fixed points of 
$U(1)^2 \times U(1)^N$ action on ${\mathcal M}_{N,k}^{\zeta}$
are in one-to-one correspondence with $N$-tuples 
of Young diagrams whose total number of boxes is equal to $k$. 
Let ${\mathcal P}_N (k)$ be the set  of such $N$-tuples of Young diagrams.
Using localization formula, we have the explicit form of 
Nekrasov's partition functions as follows;
\beq
	{\mathcal Z}^{\mathrm{inst}} (\epsilon_1 , \epsilon_2 , \vec{a} ; q) 
	= \sum_{k=0}^{\infty} \sum_{\underline{Y} \in {\mathcal P}_N (k)} 
	\frac{q^k}{\prod_{\alpha ,\beta =1}^N n_{\alpha ,\beta}^{\underline{Y}} (\epsilon_1 ,\epsilon_2 ,\vec{a})}~,
\eeq
where 
\beqa
n_{\alpha ,\beta}^{\underline{Y}} (\epsilon_1 ,\epsilon_2 ,\vec{a}) 
&:=& \prod_{s\in Y_{\alpha}} 
	(-l_{Y_{\beta}} (s) \epsilon_1 +(a_{Y_{\alpha}} (s)+1)\epsilon_2 
+ a_{\beta} -a_{\alpha}) \nonumber \\
& &~~~\times \prod_{t\in Y_{\beta}}
	((l_{Y_{\alpha}} (t)+1)\epsilon_1 -a_{Y_{\beta}} (t)\epsilon_2 
+ a_{\beta} -a_{\alpha})~,
\eeqa
$l_{Y} (s) := \nu_j -i \, , \, a_{Y} (s) := \mu_i -j$ for $s=(i,j) \in Y =(\mu_1 \geq \mu_2 \geq \cdots )$ 
and $Y^\vee= (\nu_1 \geq \nu_2 \geq \cdots)$ is the transpose of the Young diagram.

We can compute the correlation function $\langle {\mathcal O}\rangle$ using the localization formula,
if we can find an extension of the operator ${\mathcal O}$ to a $\xi$-equivariantly closed form.
In the following we set $\epsilon_1 =-\epsilon_2 =\hbar$ for simplicity.
The action of $U(1)$ on $\C^2$ is defined by $\xi_{\C^2} := i\hbar 
(z^1 \partial_{z^1} - z^2 \partial_{z^2} -h.c.)$.
We find the following forms are invariant under $\xi_{\C^2}$-action and
closed with respect to $d_{\xi_{\C^2}} := d-\iota_{\xi_{\C^2}}$, namely they are
$\xi_{\C^2}$-equivariantly closed forms \cite{FMPT},
\beq
\begin{array}{ll}
	&\alpha_{(0,0)} := 1~, \\
	&\alpha_{(2,0)} := dz^1 \wedge dz^2 +i\hbar z^1 z^2~, \\
	&\alpha_{(0,2)} := d\overline{z}^1 \wedge d\overline{z}^2 -i\hbar \overline{z}^1 \overline{z}^2~, \\
	&\alpha_{(2,2)} := \alpha_{(2,0)} \wedge \alpha_{(0,2)}~.
\end{array}
\eeq
In terms of the curvature ${\mathcal F}$ on the universal sheaf ${\mathcal E}$ 
over $\C^2 \times {\mathcal M}_{N,k}^{\zeta}$, we have the $\xi$-equivariant extension
$\trace \varphi^J \mapsto \alpha_{(2,2)} \wedge \trace {\mathcal F}^J$.
It is equivariantly closed, since ${\mathcal F} = d_{\xi} (\overline{U} d_{\xi} U)$ is exact.
Hence, the equivariant extension of scalar correlator is given by
\beq
\left\langle \trace \varphi^J \right\rangle
= \frac{1}{V{\mathcal Z}} 
			\int_{{\mathcal M}} \int_{\C^2} \alpha_{(2,2)} \wedge \trace {\mathcal F}^J
			\exp \left[-{\mathcal S}_{{\mathcal N}=1}\right]~.  \label{tracephi} \\
\eeq
These correlation functions should be regarded as equivariant integral \eqref{integral},
which one can compute by the localization formula. These equivariant integrals are
Laurent series in $\hbar$ and $a_\ell$ from which we can obtain
the original correlator \eqref{correldef} in the limit $\hbar \rightarrow 0$. 
We note the scalar correlators $\langle \trace \varphi^J \rangle$ are
independent of the superpotential $W(\varphi )$ and 
the same as ${\mathcal N}=2$ calculation \cite{FMPT}.


\section{One-point function in maximally confining phase}
\setcounter{equation}{0}

Chiral operators ${\mathcal O}$ in supersymmetric field theories are,
by definition \cite{CDSW}, annihilated by the fermionic charges 
$\overline{Q}_{\dot\alpha}$ of one chirality;
$[\overline{Q}_{\dot\alpha}, {\mathcal O}]_\pm =0$,
considered modulo $\overline{Q}_{\dot\alpha}$-exact operators;
${\mathcal O} \simeq {\mathcal O}+ [\overline{Q}_{\dot\alpha}, \Lambda]_\pm$.
From the supersymmetry algebra in four dimensions,
$[Q_\alpha, \overline{Q}_{\dot\alpha}]_+= \sigma_{\alpha \dot\alpha}^\m P_\m $, 
we can see that the correlation functions of chiral operators 
are \lq\lq topological\rq\rq\ in the sense 
that they are independent of the positions of operators. 
Especially, topological one-point functions characterize the vacuum structure
(phase) of the theory.

As we have seen in section 2, the computation of one-point function $\langle \trace \varphi^{2n} \rangle$ involves
the Chern class $\trace {\mathcal F}^{2n}$ of a universal sheaf ${\mathcal E}$ on $\C^2 \times {\mathcal M}$.
Over the instanton moduli space ${\mathcal M}_{N,k}^{\zeta}$ we have two vector bundles $W$ and $V$ of
rank $N$ and $k$, which naturally arise in the ADHM construction. 
The ADHM data are identified as $B_1, B_2 \in {\mathrm {Hom}} (V, V)$ and 
$J,  I^\dagger \in {\mathrm {Hom}} (W, V)$\footnote{In the description of ADHM construction 
in terms of $D$ branes in type IIB theory, $V$ and $W$ are the Chan-Paton bundles for $D(-1)$-branes and 
$D3$ branes, respectively.}.  Roughly speaking, the vector bundle $W$ comes from a local trivialization
of the instanton at infinity\footnote{The moduli space 
${\mathcal M}_{N,k}^{\zeta}$ is defined by the quotient by the gauge transformations that fix the  ``framing" at 
infinity.}, while $V$ is the bundle of Dirac zero modes. The fiber of $V$ is the space of 
(normalizable) solutions to the Dirac equation in the instanton background. The Riemann-Roch
theorem tells us that the number of Dirac zero modes is just the instanton number $k$. 
From vector bundles $E_1$  on $\C^2$ and $E_2$ on ${\mathcal M}_{N,k}^{\zeta}$, we can
construct an (external) tensor product bundle $E_1 \boxtimes E_2 := p_1^* E_1 \otimes p_2^* E_2$
on $\C^2 \times {\mathcal M}_{N,k}^{\zeta}$, where $p_i$ denotes the projection to the $i$-th component. 
Then as an element of the equivariant $K$-cohomology group 
the universal sheaf is isomorphic to the virtual vector bundle \cite{NY2, LMN};
\beq
{\mathcal E} \simeq  {\mathcal O}_{\C^2} \boxtimes W \oplus (S^{-} - S^{+})\boxtimes V~,
\eeq
where $S^\pm$ are positive and negative spinor bundles on $\C^2$. Their characters are
\beq
{\mathrm{Ch}(S^{+})}~(t) = 1 + e^{it(\epsilon_1 + \epsilon_2)}~, \quad
{\mathrm{Ch}(S^{-})}~(t) = e^{it\epsilon_1} + e^{it\epsilon_2}~.
\eeq
According to \cite{NY2, LMN}, at a fixed point of the toric action labeled by 
$N$-tuples of Young diagrams $Y_\alpha$
the Chern character of ${\mathcal E}$ is given by
\beq
{\mathrm{Ch}({\mathcal E})}_{\underline{Y}} (t) = \sum_{\alpha =1}^N e^{i t a_\alpha} 
- \left( 1 - e^{i t \epsilon_1} \right) \left( 1 - e^{i t \epsilon_2} \right)
\sum_{\alpha=1}^N \sum_{(k,\ell) \in Y_\alpha} e^{i t a_\alpha + i t \epsilon_1(k -1) +  i t \epsilon_2(\ell -1)}~,
\eeq
and we have
\beq
{\mathrm{Ch}}({\mathcal E}) (t) =  \sum_{\underline{Y} \in {\mathcal P}_N (k)} 
{\mathrm{Ch}({\mathcal E})}_{\underline{Y}} (t)~.
\eeq
The $n$-th Chern class $c_n({\mathcal E})$ is defined by the expansion
\beq
{\mathrm{Ch}({\mathcal E})}_{\underline{Y}} (t) 
= \sum_{n=0}^\infty \frac{(it)^{n}}{n!} c_n({\mathcal E})_{\underline{Y}}~. 
\label{chernclass}
\eeq
Since we identify ${\mathcal F}$ as a curvature on the universal bundle ${\mathcal E}$, we have
$\trace {\mathcal F}^{n} = c_{n}({\mathcal E})$.

In the following we consider $U(1)$ theory $(N=1)$ and we put $\varphi_{cl}=a=0$ for simplicity.
The fixed points are labeled by a single Young diagram $Y$ and the contribution to the character is
\beq
{\mathrm{Ch}}({\mathcal E})_Y = 1 - \left( 1 - e^{i t \epsilon_1} \right) 
\left( 1 - e^{i t \epsilon_2} \right) \sum_{(m,\ell) \in Y} 
e^{ i t \epsilon_1(m -1) +  i t \epsilon_2(\ell -1)}~. \label{chE}
\eeq
It is known that, the moduli space of instantons ${\mathcal M}_{1,k}$ for $U(1)$ case
is nothing but the Hilbert scheme of $k$-points
on $\C^2$, $(\C^2)^{[k]}$ and that 
there is a natural vector bundle ${\mathcal V}$ on $(\C^2)^{[k]}$ 
of rank $k$, called tautological vector bundle.
One can show that the (equivariant) character of ${\mathcal V}$ is the same
as the vector bundle $V$ \cite{Nak,LQW};
\beq
{\mathrm{Ch}}({\mathcal V}) = \sum_{|Y|=k} \sum_{(m,\ell) \in Y} 
e^{ i t \epsilon_1(m -1) +  i t \epsilon_2(\ell -1)}~.
\eeq
Putting $\epsilon_1=-\epsilon_2 = \hbar$ and comparing \eqref{chernclass} 
and \eqref{chE}, we find $\trace \varphi^{2n}_Y = c_{2n}({\mathcal E})_Y, (n > 0)$ is given by;
\beqa
\trace \varphi^{2n}_Y
&=& \hbar^{2n} \sum_{(k,\ell) \in Y}  \left[(k-\ell+1)^{2n} 
+ (k-\ell-1)^{2n} - 2 (k- \ell)^{2n}\right] \CR
&=& \hbar^{2n}\sum_{m=0}^{n-1} 2 \binom{2n}{2m}
\sum_{\unitbox\in Y}c(\unitbox)^{2m}~,
\label{traceformula}
\eeqa
where $c(\unitbox):=(\ell -k)$ is the content at $\unitbox=(k,\ell)$. 
On the other hand, computing geometric series, we have
\beqa
& &\sum_{(k,\ell) \in Y} (1-e^{i\epsilon_1})(1-e^{i\epsilon_2})
e^{i(k-1)\epsilon_1 + i(\ell -1) \epsilon_2} \CR
&=& \sum_{\ell=1}^{d} \sum_{k=1}^{\m_\ell} (1-e^{i\epsilon_1})(1-e^{i\epsilon_2})
e^{i(k-1)\epsilon_1 + i(\ell -1) \epsilon_2} \label{sum} \\
&=& \sum_{\ell=1}^{d} \left( e^{i(\epsilon_1 \m_\ell + \epsilon_2 \ell)}
- e^{i(\epsilon_1 \m_\ell + \epsilon_2(\ell-1))} - e^{i\epsilon_2 \ell}  
+ e^{\epsilon_2(\ell-1)}  \right)~, \nonumber
\eeqa
where $d$ is the number of rows of $Y$ and $\m_\ell$ is the number
of boxes in the $\ell$-th row. Hence, we obtain another expression of 
\eqref{traceformula};
\beq
 \trace \varphi^{2n}_Y = \hbar^{2n} \sum_{\ell=1}^{d}
 \left[ (\m_\ell -(\ell-1))^{2n} - (\m_\ell - \ell)^{2n}
- (\ell -1)^{2n} + \ell^{2n} \right]~,
\eeq
which we often find in the literature.

The partition function of $U(1)$ gauge theory is
\beq
Z_{U(1)}= \sum_{k=0}^\infty \sum_{|Y|=k} 
\frac{1}{\prod_{\unitbox \in Y} (\hbar h(\unitbox))^2} q^k~, \label{abelian}
\eeq
where $h(\unitbox)$ is the hook length at $\unitbox$ and $q=\Lambda^2$ is
the parameter of instanton expansion. The weight $\m(Y)^2=\prod_{\unitbox \in Y}(h(\unitbox))^{-2}$ 
defining $Z_{U(1)}$ is called the Plancherel measure on the space of (random) partitions.
The Plancherel measure is regarded as a discretization of 
the Vandermonde measure on random matrix. 
It is a classical result in representation theory that
\beq
\prod_{\unitbox \in Y} \frac{1}{h(\unitbox)}= \frac{{\mathrm{dim}}~S^Y}{k!}~, \label{dimension}
\eeq
where $S^Y$ is the irreducible representation of the symmetric group labeled by
a Young diagram $Y$.
By the Plancherel formula $\sum_{|Y|=k} ({\mathrm{dim}}~S^Y)^2 = k!$, we obtain
\beq
\sum_{|Y|=k} \prod_{\unitbox \in Y} h(\unitbox)^{-2} = \frac{1}{k!}~.
\label{norm}
\eeq
Hence we find that the summation over the instanton number $k$ in \eqref{abelian} is 
organized into a simple form \cite{LMN};
\beq
Z_{U(1)} = \exp \left( \frac{q}{\hbar^2} \right)~.
\eeq

The correlation functions of our interest are
\beq 
\langle \trace \varphi^{2n} \rangle = \frac{1}{Z_{U(1)}} 
\sum_{k=1}^\infty \sum_{|Y|=k} 
\frac{\trace~\varphi^{2n}_Y}
{\hbar^{2k} \prod_{\unitbox \in Y} h(\unitbox)^2} q^k~. \label{correlator}
\eeq
Substituting the formula \eqref{traceformula}, we have
\beq
\langle \trace \varphi^{2n} \rangle 
\exp \left( \frac{q}{\hbar^2} \right) =  2 \sum_{m=0}^{n-1} \binom{2n}{2m} 
\sum_{k=1}^\infty S_m(k) \hbar^{2(n-k)} q^k~,
\eeq
where we have introduced
\beq
S_n(k) := \sum_{|Y|=k} \frac{\sum_{\unitbox\in Y} c(\unitbox)^{2n}} 
{\prod_{\unitbox \in Y} h(\unitbox)^2}~. \label{Sndef}
\eeq
Thus the computation of $\langle \trace \varphi^{2n} \rangle$
is equivalent to giving summation formula for $S_n(k)$ over Young diagrams.
For example, a  ``trivial" formula $S_0(k)= 1/(k-1)!$ implies
 $\langle \trace\varphi^{2} \rangle =2q$.
Looking at the instanton expansion of lower degree explicitly, we find
\beqa
(6q^2 + 2 \hbar^2 q) e^{\frac{q}{\hbar^2}}
&=& \sum_{k=1}^\infty \left( 2 S_0(k) + 12 S_2(k) \right) 
\hbar^{4-2k}q^k ~, \CR
(20 q^3 + 30 \hbar^2 q^2 + 2 \hbar^4 q) e^{\frac{q}{\hbar^2}}
&=& \sum_{k=1}^\infty \left( 2 S_0(k) + 30 S_2(k) + 30 S_4(k) \right) 
\hbar^{6-2k} q^k~, \label{example} \\
(70 q^4+280 \hbar^2 q^3+126 \hbar^4 q^2+ 2 \hbar^6 q) e^{\frac{q}{\hbar^2}}
&=& \sum_{k=1}^\infty \left( 2 S_0(k) + 56 S_2(k) + 140 S_4(k) + 56 S_6(k) \right) 
\hbar^{8-2k} q^k~. \nonumber
\eeqa
In Appendix we prove the following formula;
\begin{align}
\sum_{j=1}^n c_j^n S_j(k)=\frac{(2n)!}{((n+1)!)^2}\frac{1}{(k-n-1)!}~,
\label{combi}
\end{align}
where $c_j^n$ are defined by\footnote{The functions $x^{\overline n}$ and $x^{\underline n}$ are
natural power functions in the calculus of difference.}
\begin{align}
{\mathcal P}_{2n}(x) := x^{\overline n} \cdot x^{\underline n} 
= \prod_{j=0}^{n-1}(x^2-j^2)=\sum_{j=1}^n c_j^n x^{2j}~.
\label{defc}
\end{align}
We note that in terms of a specialization of the elementary 
symmetric functions $e_r(x)$, the coefficient $c_j^n$ is 
given by
\begin{align}
c_j^n = (-1)^{n-j} e_{n-j}(1^2, 2^2, \cdots, (n-1)^2)~.
\end{align}
The formula implies, for example, 
\beqa
& &S_2(k) = \frac{1}{2} \frac{1}{(k-2)!}~, 
\qquad
S_4(k) = \frac{1}{2} \frac{1}{(k-2)!} + \frac{2}{3} \frac{1}{(k-3)!}~, \CR
& & S_6(k) = \frac{1}{2} \frac{1}{(k-2)!} + \frac{10}{3} \frac{1}{(k-3)!} 
+ \frac{5}{4} \frac{1}{(k-4)!}~,
\eeqa
and we find an agreement with \eqref{example}.

In general, based on the combinatorial formula \eqref{combi},
we can derive the following relation among topological one-point functions valid in the 
maximally confining phase;
\begin{align}
\sum_{j=1}^rc_j^r\hbar^{2(r-j)}\langle\trace\varphi^{2j}\rangle
=\frac{(2r)!}{(r!)^2}q^r~.
\label{noquantum}
\end{align}
In other words, the linear combination on the left hand side with extra
terms of the coefficients $c_j^r$ is a ``good'' combination without
quantum corrections. Before embarking a proof of \eqref{noquantum},
let us take a look at some examples first.
From \eqref{example} it is easy to find that\footnote{These one-point functions 
in the limit $\hbar \to 0$ were computed in \cite{FMPT}.}
\begin{align}
\langle\trace\varphi^2\rangle&=2q~, \CR
\langle\trace(\varphi^4-\hbar^2\varphi^2)\rangle&=6q^2~,  \CR
\langle\trace(\varphi^6-5\hbar^2\varphi^4+4\hbar^4\varphi^2)\rangle
&=20q^3~, \\
\langle\trace(\varphi^8-14\hbar^2\varphi^6
+49\hbar^4\varphi^4-36\hbar^6\varphi^2)\rangle&=70q^4~, \nonumber
\end{align}
and we recognize the coefficients $c^r_j$ in the linear combinations of 
$\langle\trace\varphi^{2n} \rangle$.

For the proof of \eqref{noquantum}, we first plug the formula \eqref{traceformula}
into the definition \eqref{correlator} of $\langle\trace\varphi^{2j}\rangle$ to obtain
\begin{align}
&\sum_{j=1}^rc_j^r\hbar^{2(r-j)}\langle\trace\varphi^{2j}\rangle Z_{U(1)}
\nonumber\\
&=\hbar^{2r}\sum_{k=1}^\infty\sum_{|Y|=k}
\biggl(\frac{q}{\hbar^2}\biggr)^k
\sum_{j=1}^rc_j^r\frac{\sum_{\unitbox\in Y}\bigl(c(\unitbox)+1\bigr)^{2j}
-2\bigl(c(\unitbox)\bigr)^{2j}+\bigl(c(\unitbox)-1\bigr)^{2j}}
{\prod_{\unitbox\in Y}h(\unitbox)^2} \CR
&=\hbar^{2r}\sum_{k=1}^\infty\sum_{|Y|=k}
\biggl(\frac{q}{\hbar^2}\biggr)^k \frac{\sum_{\unitbox\in Y}\Bigl[
{\mathcal P}_{2r}(c(\unitbox)+1) -2 {\mathcal P}_{2r}(c(\unitbox))
+ {\mathcal P}_{2r}(c(\unitbox)-1)
\Bigr]}{\prod_{\unitbox\in Y}h(\unitbox)^2} \label{proof} \\
&= 2r(2r-1)\hbar^{2r}\sum_{k=1}^\infty\sum_{|Y|=k}
\biggl(\frac{q}{\hbar^2}\biggr)^k \frac{\sum_{\unitbox\in Y}
\prod_{j=0}^{r-2}\bigl(c(\unitbox)^2-j^2\bigr)}
{\prod_{\unitbox\in Y}h(\unitbox)^2}~, \nonumber
\end{align}
where in the last line we have used
the following relation satisfied by ${\mathcal P}_{2n}(x)$\footnote{
This formula is a discrete version of $\frac{d^2}{dx^2} x^{2n} = 2n(2n-1) x^{2n-2}$.};
\beq
\Delta^2 {\mathcal P}_{2n}(x) := {\mathcal P}_{2n}(x+1) -2 {\mathcal P}_{2n}(x)
+ {\mathcal P}_{2n}(x-1) = 2n(2n-1) {\mathcal P}_{2n-2}(x)~.
\eeq
Finally as is shown in Appendix, the formula \eqref{combi} is equivalent to 
\begin{align}
\sum_{|Y|=k}
\frac{\sum_{\unitbox\in Y}\prod_{j=0}^{r-2}\bigl(c(\unitbox)^2-j^2\bigr)}
{\prod_{\unitbox\in Y}h(\unitbox)^2}
=\frac{(2(r-1))!}{(r!)^2}\frac{1}{(k-r)!}~,
\end{align}
which allows us to factorize the partition function $Z_{U(1)}$ as follows;
\begin{align}
\sum_{j=1}^rc_j^r\hbar^{2(r-j)}\langle\trace\varphi^{2j}\rangle Z_{U(1)}
=\frac{(2r)!}{(r!)^2}q^r\sum_{k=1}^\infty
\frac{1}{(k-r)!}\biggl(\frac{q}{\hbar^2}\biggr)^{k-r}
=\frac{(2r)!}{(r!)^2}q^r\exp\biggl(\frac{q}{\hbar^2}\biggr)~.
\end{align}
Dividing both sides by the partition function $Z_{U(1)}$, we obtain \eqref{noquantum}.

\section{Genus Expansion}
\setcounter{equation}{0}

From the relation \eqref{noquantum}
derived in section 3, we can compute the expansion of the generating function
\beq
T(z) := \left\langle \trace \frac{1}{z-\varphi} \right\rangle
= \sum_{n=0}^\infty z^{-n-1} \langle \trace \varphi^n \rangle~, 
\eeq
in $\hbar^2$ iteratively. The expansion should 
be compared with the genus expansion of topological strings
and/or matrix models. 
Recall that the coefficients $c_j^r$ are defined by
\beq
\prod_{j=0}^{r-1} (x^2 - j^2) = \sum_{j=1}^r c_j^r x^{2j}~.
\eeq
We find
\beq
c_r^r = 1~, \qquad c_{r-1}^r = - \sum_{j=0}^{r-1} j^2 
= - \frac{1}{6} r (r-1) (2r-1)~.
\eeq
Substituting these to the relation \eqref{noquantum}, we obtain
\beqa
\langle \trace \varphi^{2r} \rangle &=& \frac{(2r)!}{(r!)^2} q^r
+ \frac{\hbar^2 }{6} r (r-1) (2r-1) \frac{(2r-2)!}{((r-1)!)^2} q^{r-1}
+ O(\hbar^4) \CR
&=& \frac{(2r)!}{(r!)^2} q^r + \frac{\hbar^2 }{12} 
\frac{(2r)!}{(r-1)! (r-2)!} q^{r-1} + O(\hbar^4)~.
\eeqa
Hence
\beq
T(z) = \sum_{n=0}^\infty z^{-2n-1} \frac{(2n)!}{(n!)^2} q^n
+ \sum_{n=2}^\infty  z^{-2n-1} \frac{\hbar^2 }{12} 
\frac{(2n)!}{(n-1)! (n-2)!} q^{n-1} + O(\hbar^4)~.
\eeq
We note that the Taylor expansion;
\beq
\frac{1}{\sqrt{1-4x}} = \sum_{n=0}^\infty 
\frac{(2n)!}{(n!)^2} x^n~, \qquad |x| < \frac{1}{4}~, \label{taylor}
\eeq
implies
\beq
T(z) = \frac{1}{z\sqrt{1- \frac{4q}{z^2}}} = \frac{1}{\sqrt{z^2 -4q}}~,
\qquad \hbar \to 0~.
\eeq
Combining the expansion \eqref{taylor} and its derivatives, 
we find
\beq
\frac{24x(x+1)}{(1-4x)^{\frac{7}{2}}} 
= \sum_{n=2}^\infty \frac{(2n)!}{(n-1)! (n-2)!} x^{n-1}~, \label{taylor2}
\eeq
which implies the generating function $T(z)$ up to genus one;
\beqa
T(z) &=& \frac{1}{\sqrt{z^2 -4q}} + \hbar^2
\frac{2q(q+z^2)}{(z^2-4q)^{\frac{7}{2}}} + O(\hbar^4) \CR
&=& \frac{1}{\sqrt{z^2 -4q}}
\left( 1 + \hbar^2\frac{2q(q+z^2)}{(z^2-4q)^3}  
+ O(\hbar^4) \right)~.
\eeqa

Similarly, the relation \eqref{noquantum} implies
the genus two part of $\langle \trace \varphi^{2r}\rangle$ is
\beq
\hbar^4 (c^r_{r-1} c^{r-1}_{r-2} - c^r_{r-2}) \frac{(2r-4)!}{((r-2)!)^2} q^{r-2}~.
\eeq
From the definition of $c_r^j$ we find
\beq
c_{r-1}^r c_{r-2}^{r-1} = \frac{1}{36} r (r-1) (2r-1)(r-2)(r-1)(2r-3)~,
\eeq
and
\beqa
c^r_{r-2} &=& \sum_{i=1}^{r-2} \sum_{j=i+1}^{r-1} i^2 j^2
= c_{r-1}^r c_{r-2}^{r-1}  - \frac{1}{6} \sum_{i=1}^{r-2} i^3(i+1)(2i+1) \CR
&=& c_{r-1}^r c_{r-2}^{r-1} -\frac{1}{360} r(5r-11)(r-1)(r-2)(2r-1)(2r-3)~.
\eeqa
Hence the genus two part of the generating function $T(z)$
is given by
\beq
\frac{\hbar^4}{1440} z^{-5} \sum_{n=3}^\infty
(5n-11) \frac{(2n)!}{(n-2)!(n-3)!} \left(\frac{q}{z^2}\right)^{n-2}~.
\eeq
From the Taylor expansion \eqref{taylor2} which we have used for
genus one part, we further obtain
\beq
\frac{2880x(27x^3+118x^2+37x+1)}{(1-4x)^{\frac{13}{2}}} 
= \sum_{n=3}^\infty (5n-11) \frac{(2n)!}{(n-2)! (n-3)!} x^{n-2}~.
\eeq
In summary the generating function up to genus two is
\beq
T(z) = \frac{1}{\sqrt{z^2 -4q}}
\left( 1 + \hbar^2\frac{2q(q+z^2)}{(z^2-4q)^3}  
+ \hbar^4 \frac{2q(27q^3 + 118 q^2 z^2 + 37 q z^4 + z^6)}{(z^2-4q)^6} 
+O(\hbar^6) \right)~.
\eeq

\section{Computations in operator formalism}
\setcounter{equation}{0}

Due to the correspondence of Young diagrams (or Maya diagrams) and
fermion Fock states with neutral charge,
operator formalism is very powerful for computations of
summations over functions on the set of Young diagrams.
Let us introduce a pair of charged (NS) free fermions
\beq
\psi(z) = \sum_{r \in {\mathbb Z} + \frac{1}{2}} \psi_r z^{-r-\frac{1}{2}}~,
\quad \psi^*(z) = \sum_{s \in {\mathbb Z} + \frac{1}{2}} \psi^*_s z^{-s-\frac{1}{2}}~,
\eeq
with the anti-commutation relation
\beq
\{ \psi_r, \psi_s^* \} = \delta_{r+s,0}~, \quad r,s \in {\mathbb Z} + \frac{1}{2}~. \label{ACR}
\eeq
The Fock vacuum $|0\rangle$ is defined by
\beq
\psi_r |0\rangle = \psi^*_s |0\rangle =0~, \quad r, s > 0~.
\eeq
Using Young/Maya diagram correspondence, for each
partition $\lambda$, we have a state $|\lambda\rangle$ 
in the charge zero sector of the fermion Fock space, 
which is given by 
\beq
|\lambda \rangle = \prod_{i=1}^\infty \psi_{i-\lambda_i -\frac{1}{2}}
 |\!| 0 \rangle\!\rangle~,
\eeq
with
\beq
\psi_s^*|\!| 0 \rangle\!\rangle =0~,\quad \forall s~.
\eeq
Recall the standard bosonization rule;
\beqa
& & J(z) = : \psi(z) \psi^*(z) : = \sum_{n \in {\mathbb Z}} J_n z^{-n-1}~, \quad
J_n = \sum_{r \in {\mathbb Z} + \frac{1}{2}} : \psi_r  \psi^*_{n-r}:~, \CR
& &J(z) = i \partial \phi (z)~, \quad \psi(z) = : e^{i\phi(z)}:~, \quad \psi^*(z) = : e^{-i\phi(z)}:~,
\eeqa
where $: \quad :$ means the normal ordering. 
Now a crucial point is the following formula
\beq
\exp\left(\frac{J_{-1}}{\hbar}\right) | 0\rangle = \sum_{k=0}^\infty \frac{1}{\hbar^k}
\sum_{|\lambda| =k} \frac{1}{\prod_{\unitbox \in \lambda} h(\unitbox)}
| \lambda \rangle~, \label{NO}
\eeq
which is eq.(5.29) of \cite{NO}.
In the language of symmetric functions, the corresponding formula is
given in \cite{Mac}. 

It is instructive to compute $J_{-1}  | 0\rangle, J_{-1}^2  | 0\rangle, J_{-1}^3  | 0\rangle, J_{-1}^4  | 0\rangle \cdots$,
iteratively. One can recognize that the action of
\begin{align}
J_{-1} =\quad&\psi_{-\frac{1}{2}}  \psi^*_{-\frac{1}{2}}  \CR
+\,& \psi_{-\frac{3}{2}}  \psi^*_{\frac{1}{2}} +  \psi_{-\frac{5}{2}}  \psi^*_{\frac{3}{2}}  
+ \psi_{-\frac{7}{2}}  \psi^*_{\frac{5}{2}} + \cdots \CR
-\,& \psi^*_{-\frac{3}{2}}  \psi_{\frac{1}{2}} -  \psi^*_{-\frac{5}{2}}  \psi_{\frac{3}{2}}  
- \psi^*_{-\frac{7}{2}}  \psi_{\frac{5}{2}} + \cdots~,
\end{align}
on the fermion Fock states is to move  ``black ball" to the right by one unit
whenever possible, if the vacuum is identified as the Maya diagram whose negative positions are
completely filled with  ``black balls". The combinatorics of this procedure gives 
\beq
J_{-1}^k |0\rangle =\sum_{|\lambda| =k} 
\frac{k!}{\prod_{\unitbox \in \lambda}h(\unitbox)} |\lambda \rangle
 =\sum_{|\lambda| =k} (\dim S^\lambda)|\lambda \rangle~. \label{J-1power}
\eeq
It is easy to compute
\beq
\langle 0| e^{\frac{J_1}{\hbar}} e^{\frac{J_{-1}}{\hbar}} | 0 \rangle 
= \langle 0| e^{\frac{[J_1,J_{-1}]}{\hbar^2}} | 0 \rangle
= e^{\frac{1}{\hbar^2}} = \sum_{k=0}^\infty \frac{1}{\hbar^{2k} k!}~.
\eeq
On the other hand
\beq
\langle 0| e^{\frac{J_1}{\hbar}} e^{\frac{J_{-1}}{\hbar}} | 0 \rangle 
= \sum_{\mu, \lambda} \langle \mu| 
\frac{1}{\hbar^{|\mu|} \prod_{\unitbox \in \mu} h(\unitbox)} \cdot
 \frac{1}{\hbar^{|\lambda|} \prod_{\unitbox \in \lambda} h(\unitbox)}
| \lambda \rangle 
=\sum_\lambda  \frac{1}{\hbar^{2|\lambda|}\prod_{\unitbox \in \lambda} h(\unitbox)^2}~,
\eeq
where we have used $\langle \mu | \lambda \rangle = \delta_{\mu, \lambda}$.
Comparing the coefficients of $\hbar^{-2k}$ of both sides, we recover the identity \eqref{norm}.
Recall that what we want to compute is $S_n(k)$ defined by \eqref{Sndef}.
Let us introduce the generating function of $S_n(k)$;
\beq
\mathrm{Ch}[k] (z) := \sum_{|\lambda|=k} \frac{\sum_{\unitbox\in \lambda} 
\exp \left(z c(\unitbox)\right)} {\prod_{\unitbox \in \lambda} h(\unitbox)^2}
= \sum_{n=0}^\infty \frac{z^{2n}}{(2n)!} S_n(k)~, \label{SnGen}
\eeq
where we have used the fact that $c(\unitbox)$ is odd under the transpose of the Young diagram.
This generating function gives the Chern character of the tautological vector
bundle over $(\C^2)^{[k]}$ considered in \cite{LQW}. 
The sum in the numerator is 
\beq
\sum_{\unitbox\in \lambda} \exp \left(z c(\unitbox)\right)
= \sum_{j=1}^{d(\lambda)} \sum_{i=1}^{\lambda_j} e^{z(i-j)}
= \sum_{j=1}^{d(\lambda)} 
\frac{e^{z(\lambda_j -j +\frac{1}{2})}- e^{z(-j+\frac{1}{2})}}
{e^{\frac{z}{2}} - e^{-\frac{z}{2}}}~.
\eeq

Following Okounkov and Pandharipande \cite{OP1,OP2}, we consider the operator 
\beq
{\mathcal E}_n(z) := \sum_{r \in {\mathbb Z} + \frac{1}{2}} e^{z\left( r - \frac{n}{2}\right)} E_{n-r,r}~,
\label{OPop}
\eeq
where $E_{r,s} := :\psi_r \psi_s^*:$ is
the standard basis of ${\mathfrak {gl}}(\infty)$ acting on the fermion Fock space\footnote{
Originally the definition in \cite{OP1,OP2} has the constant term $\frac{\delta_{n,0}}{{\mathrm{sh}}(z)}$, 
which eliminates the central extension term in the commutation relation \eqref{Wcom}. 
Note also that our convention of the anti-commutation relation \eqref{ACR}
is different from the original one in \cite{OP1,OP2}, where the right hand side is $\delta_{r,s}$. }.
We can see that ${\mathcal E}_n(z)$ satisfies the commutation relation
\beq
[ {\mathcal E}_n(z), {\mathcal E}_m(w)] = {\mathrm{sh}}(nw-mz)~{\mathcal E}_{n+m}(z+w)
+ \delta_{n+m,0} \frac{{\mathrm{sh}}\,n(z+w)}{{\mathrm{sh}}\,(z+w)}~, \label{Wcom}
\eeq
where ${\mathrm{sh}}(z):= e^{\frac{z}{2}} - e^{-\frac{z}{2}}$. We have ${\mathcal E}_n(0) = J_n$ with
$J_n$ being the modes of the standard $U(1)$ current of fermions.
We also find an important relation
\beqa
{\mathcal E}_0(z) |\lambda \rangle 
&=& \sum_{i=1}^\infty \left( 
 e^{z\left(\lambda_i -i +\frac{1}{2} \right)} 
- e^{z\left(\frac{1}{2} -i\right)}  \right) |\lambda \rangle
= \sum_{i=1}^{d(\lambda)} \left( 
 e^{z\left(\lambda_i -i +\frac{1}{2} \right)} 
- e^{z\left(\frac{1}{2} -i \right)}  \right) |\lambda \rangle \CR
&=& {{\mathrm{sh}}(z)} \sum_{\unitbox\in \lambda} 
\exp \left(z c(\unitbox)\right)  |\lambda \rangle~.
\eeqa
The second term comes from ${\cal E}_0(z) |\!|0 \rangle\!\rangle
= -({\mathrm{sh}} (z))^{-1} |\!|0 \rangle\!\rangle$, which can be
calculated directly from the definition of $|\!|0\rangle\!\rangle$ 
or the consistency ${\cal E}_0(z)|0\rangle=0$.
By the formula \eqref{J-1power}, the generating function of $S_n(k)$ is 
expressed in operator formalism as follows;
\beq
(k!)^2 {\mathrm{sh}}(z) \mathrm{Ch}[k](z) = \langle 0| J_1^k {\mathcal E}_0(z) J_{-1}^k|0\rangle~.
\eeq
The right hand side can be computed by the commutation relation \eqref{Wcom}, which implies
\beq
J_1^k {\mathcal E}_0(z) = \sum_{\ell=0}^k 
\binom{k}{\ell} {\mathrm{sh}}^\ell (z) {\mathcal E}_\ell (z) J_1^{k-\ell}~.
\eeq
We also use
\beq
J_1^n J_{-1}^k |0 \rangle = \frac{k!}{(k-n)!} J_{-1}^{k-n} |0 \rangle~, \quad (n \leq k)~,
\eeq
which is derived from $e^{zJ_1} e^{wJ_{-1}} |0 \rangle = 
e^{[zJ_1, wJ_{-1}]} e^{wJ_{-1}}  e^{zJ_1} |0 \rangle = e^{zw} e^{wJ_{-1}}|0 \rangle$.
By these formulae, we obtain
\beq
\langle 0| J_1^k {\mathcal E}_0(z) J_{-1}^k|0\rangle
= \sum_{\ell=0}^k \binom{k}{\ell}
{\mathrm{sh}}^\ell (z) \langle 0 | {\mathcal E}_\ell (z) \frac{k!}{\ell!} J_{-1}^\ell |0 \rangle
= (k!)^2 \sum_{\ell=1}^k \frac{{\mathrm{sh}}(z)^{2\ell-1}}{(\ell!)^2 (k-\ell)!}~.
\eeq
The contribution from $\ell=0$ is simply zero because $\langle 0|{\cal E}_0(z)|0\rangle=0$.
Thus we find\footnote{From the Lascoux-Thibon formula used in Appendix, we see that
$\mathrm{Ch}[k] (z)= \sum_{m=1}^{k} \phi_{(1^m)}(z) \frac{1}{(k-m)!}$.}
\beq
\mathrm{Ch}[k] (z) = \sum_{\ell=1}^k \frac{{\mathrm{sh}}(z)^{2\ell-2}}
{(\ell !)^2 (k-\ell)!}~. \label{character}
\eeq
We note that the constant term of \eqref{character} gives
\beq
\mathrm{Ch}[k](0) = \frac{1}{(k-1)!}~,
\eeq
which is consistent with
\beq
S_0(k) = \frac{k}{k!}~,
\eeq
derived from \eqref{norm}. We have computed
the Taylor expansion of $\mathrm{Ch}[k] (z)$ for each fixed $k$ and
found exact agreements with the results of the formula \eqref{combi} proved
in Appendix. Note that the formula \eqref{combi} rather gives $S_n(k)$ 
as a function of $k$ for each fixed $n$. 


\section{Loop operator}
\setcounter{equation}{0}

In this section\footnote{We thank Amihay Hanany and Hiroyuki
Ochiai for sharing the ideas that led us to make the following computations.
}, let us consider the vacuum expectation value of the
loop operator 
\begin{align}
\langle\trace e^{it\varphi}\rangle
=\sum_{n=0}^\infty\frac{(it)^n}{n!}\langle \trace \varphi^n \rangle~.
\label{wilson}
\end{align}
Note that the loop operator is related to the resolvent operator
simply by the Laplace transformation:
\begin{align}
T(z)=\biggl\langle \trace \frac{1}{z-\varphi}\biggr\rangle
=\int_0^\infty dle^{-lz}\langle\trace e^{l\varphi}\rangle~. \label{laplace}
\end{align}

Plugging \eqref{correlator} with \eqref{traceformula} into \eqref{wilson}, 
we find the loop operator $\langle\trace e^{it\varphi}\rangle$ can be expressed as:
($z=it\hbar$)
\begin{align}
\langle\trace e^{it\varphi}\rangle -1 =\frac{1}{Z_{U(1)}}
\sum_{k=1}^\infty\biggl[\frac{q}{\hbar^2}\biggr]^k \sum_{|Y|=k} 
\frac{\sum_{\unitbox\in Y}(e^{z(c(\unitbox)+1)}+e^{z(c(\unitbox)-1)}
-2e^{z(c(\unitbox))})} {\prod_{\unitbox\in Y}(h(\unitbox))^2}~,
\end{align}
which can further be put into
\begin{align}
\langle\trace e^{it\varphi}\rangle -1
&=\frac{1}{Z_{U(1)}}
\sum_{k=1}^\infty\biggl[\frac{q}{\hbar^2}\biggr]^k
{\mathrm{sh}}^2(z) {\mathrm{Ch}}[k](z)~,
\label{wilsonch}
\end{align}
using the function ${\mathrm{Ch}}[k](z)$ defined in \eqref{SnGen}.
Since we have already evaluated ${\mathrm{Ch}}[k](z)$, let us plug the final
expression of $ {\mathrm{Ch}}[k](z)$ in \eqref{character} into \eqref{wilsonch}:
\begin{align}
\langle\trace e^{it\varphi}\rangle -1
=\frac{1}{Z_{U(1)}}\sum_{k=1}^\infty\sum_{\ell=1}^k
\biggl[\frac{q}{\hbar^2}\biggr]^k\frac{{\mathrm{sh}}(z)^{2\ell}}{(\ell!)^2 (k-\ell)!}
=\frac{1}{Z_{U(1)}}\sum_{\ell=1}^\infty\sum_{k=\ell}^\infty
\biggl[\frac{q}{\hbar^2}\biggr]^k\frac{{\mathrm{sh}}(z)^{2\ell}}{(\ell!)^2(k-\ell)!}~,
\end{align}
where in the last equation we have exchanged the $k$ summation and the
$\ell$ summation.
If we perform the $k$ summation first,
\begin{align}
\sum_{k=\ell}^\infty\frac{1}{(k-\ell)!}\biggl[\frac{q}{\hbar^2}\biggr]^k
=\sum_{k=0}^\infty\frac{1}{k!}\biggl[\frac{q}{\hbar^2}\biggr]^{k+\ell}
=Z_{U(1)}\biggl[\frac{q}{\hbar^2}\biggr]^\ell~,
\end{align}
we find finally the loop operator is given as
\begin{align}
\langle\trace e^{it\varphi}\rangle
= 1+ \sum_{\ell=1}^\infty\frac{1}{(\ell!)^2}
\biggl[\frac{q}{\hbar^2} {\mathrm{sh}}^2(z) \biggr]^\ell
=I_0\bigl(2\sqrt{q}\,{\mathrm{sh}}(it\hbar)/\hbar\bigr)~, \label{bessel}
\end{align}
with $I_n(x)$ being the modified Bessel functions. It is interesting that we can perform
the instanton sum of the loop operator in a closed form and obtain 
the exact result \eqref{bessel} in the parameter $\hbar$. 

Finally we can also obtain an exact result on $T(z)$ from \eqref{bessel}. 
The Laplace transformation \eqref{laplace} implies
\beqa
T(z) &=& \sum_{n=0}^\infty \frac{1}{(n!)^2}
\left( \frac{q}{\hbar^2} \right)^n \int_0^\infty dl e^{-lz} 
{\mathrm{sh}}^{2n} (l \hbar) \CR
&=& \sum_{n=0}^\infty \frac{(2n)!}{(n!)^2} \left( \frac{q}{\hbar^2} \right)^n
\sum_{m=-n}^n \frac{(-1)^{n-m}}{(n-m)!(n+m)!} \frac{1}{z - m\hbar}~.
\eeqa
By computing the residue at $z=m\hbar~(-n \leq m \leq n)$, we have
the partial fraction expansion\footnote{We note that the same formula has been used
in the proof of (A.3).}
\beq
\prod_{m=-n}^n \frac{1}{z-m\hbar} 
= \hbar^{-2n} \sum_{m=-n}^n  \frac{(-1)^{n-m}}{(n-m)!(n+m)!}\frac{1}{z - m\hbar}~,
\eeq
which gives
\beq
T(z) = \sum_{n=0}^\infty \frac{(2n)!}{(n!)^2} q^n 
\prod_{m=-n}^n \frac{1}{z-m\hbar} 
= \sum_{n=0}^\infty \frac{(2n)!}{(n!)^2} q^n z^{-2n-1}
\prod_{m=1}^n \sum_{k=0}^\infty \left(\frac{m\hbar}{z}\right)^{2k}~. \label{complete}
\eeq
This is a rather simple answer to the $\hbar$ dependent of $T(z)$,
which is consistent with the $\hbar$ expansion (up to genus two)
presented in section 4.


\section*{Acknowledgements}

We would like to thank Hiraku Nakajima for a helpful correspondence.
We are grateful to Satoshi Minabe and Hiroyuki Ochiai 
for discussions and comments on the manuscript. 
We also thank Freddy Cachazo and Amihay Hanany for discussions.
Part of the results in this paper was presented in MSJ-IHES joint workshop on Non-commutativity,
held at IHES (Bures-sur-Yvette) in November 2006. One of the authors (H.K.) 
would like to thank the organizers for the invitation. The work of S.M. was supported
partly by Inamori Foundation, Nishina Memorial Foundation and Grant-in-Aid for 
Young Scientists (\# 18740143) from the Japan Ministry of Education, Culture,
Sports, Science and Technology.


\newtheorem{theorem}{Theorem}[section]
\newtheorem{prop}[theorem]{Proposition}
\newtheorem{lemma}[theorem]{Lemma}
\newtheorem{corollary}[theorem]{Corollary}
\newtheorem{definition}[theorem]{Definition}
\newtheorem{example}[theorem]{Example}
\newtheorem{remark}{Remark}
\newtheorem{conjecture}[theorem]{Conjecture}
\newtheorem{problem}[theorem]{Problem}

\newenvironment{demo}[1]{%
  \trivlist
  \item[\hskip\labelsep
        {\bf #1.}]
}{%
  \endtrivlist
}

\newcommand\Comp{\mathbb{C}}
\newcommand\Sym{\mathfrak{S}}
\newcommand\ep{\varepsilon}


\setcounter{section}{1}
\renewcommand{\thesection}{\Alph{section}}

\section*{Appendix: Proof of the Combinatorial Identity}
\setcounter{equation}{0}

The aim of this appendix is to give a proof to the following theorem.

\begin{theorem} \label{thm:a1}
\begin{equation}
\sum_{\lambda \vdash k}
 \frac{ \sum_{x \in \lambda} \prod_{i=0}^{r-1} (c(x)^2 - i^2) }
      { \prod_{x \in \lambda} h(x)^2 }
 =
\frac{ (2r)! }{ ( (r+1)! )^2 }
\cdot
\frac{ \prod_{i=0}^r (k-i) }
     { k! },
\label{a1}
\end{equation}
\end{theorem}
where $\lambda$ runs over all partitions of $k$, i.e., the Young diagrams with $k$ boxes.

We put
$$
S_r(k)
 =
\sum_{\lambda \vdash k}
 \frac{ \sum_{x \in \lambda} c(x)^{2r} }
      { \prod_{x \in \lambda} h(x)^2 },
\quad
T_r (k)
 =
\sum_{\lambda \vdash k}
 \frac{ \sum_{x \in \lambda} \prod_{i=0}^{r-1} (c(x)^2-i^2) }
      { \prod_{x \in \lambda} h(x)^2 },
$$
and denote by $e_i$ and $h_i$ the $i$-th elementary and complete symmetric
 polynomial respectively.
Then we have
$$
T_r(k)
 =
 \sum_{p=1}^r (-1)^{r-p} e_{r-p}( 1^2, 2^2, \cdots, (r-1)^2 ) S_p(k).
$$
Since the matrices
$$
\left( (-1)^{j-i} e_{j-i}(x_1, \cdots, x_j) \right)_{0 \le i, j \le N}
\quad
\text{and}
\quad
\left( h_{j-i} (x_1, \cdots, x_{i+1}) \right)_{0 \le i, j \le N}
$$
are inverses to each other, we see that
$$
S_r(k)
 =
  \sum_{p=1}^r h_{r-p}( 1^2, 2^2, \cdots, p^2 ) T_p(k).
$$
Thus the identity (\ref{a1}) is equivalent to
\begin{equation}
S_r(k)
 =
\sum_{p=1}^r h_{r-p}(1^2,2^2,\cdots,p^2)
 \frac{(2p)!}{((p+1)!)^2} \frac{ k^{\underline{p+1}} }{ k! },
\label{a2}
\end{equation}
where $k^{\underline{p+1}}$ denotes the falling factorial
$$
k^{\underline{p+1}} = k (k-1) \cdots (k-p).
$$

We prove the identity (\ref{a2}) by using the Jucys--Murphy elements $L_i$
 ($1 \le i \le k$) in the group ring $\Comp[\Sym_k]$ of the symmetric group $\Sym_k$
 and the Lascoux--Thibon formula for them.
The Jucys--Murphy elements (\cite{J,Mur})
are defined to be the sum of
 transpositions
$$
L_i = (1,i) + (2,i) + \cdots + (i-1,i).
$$
Note that $L_1 = 0$, but it is convenient to include this case.
A key property of Jucys--Murphy elements is the following.

\begin{prop} \label{prop:a2}
\begin{enumerate}
\item
The Jucys--Murphy elements $L_1, \cdots, L_k$ are commutative.
\item
On the irreducible representation $S^\lambda$ of $\Sym_k$
 corresponding to a partition $\lambda$,
 the operators $L_1, \cdots, L_k$ are simultaneously diagonalizable
 and the eigenvalues of $L_i$ are the contents $\{ c(x) : 
x \in \lambda \}$ of $\lambda$.
\end{enumerate}
\end{prop}

\begin{prop} \label{prop:a3}
Let $f(z)$ be a polynomial.
Then the quantity
$$
k!
\sum_{\lambda \vdash k}
 \frac{ \sum_{x \in \lambda} f(c(x)) }
      { \prod_{x \in \lambda} h(x)^2 }
$$
is equal to the coefficient of the identity element in
 $f(L_1) + \cdots + f(L_k)$.
\end{prop}

\begin{demo}{Proof}
Since $f(L_1) + \cdots + f(L_k)$ is symmetric in $L_1, \cdots, L_k$,
 we see that $f(L_1) + \cdots + f(L_k)$ acts on $S^\lambda$
 as the scalar multiplication by $\sum_{x \in \lambda} f(c(x))$.
Hence the trace of the operator $f(L_1) + \cdots + f(L_k)$ on $S^\lambda$
 is equal to $f^\lambda \sum_{x \in \lambda} f(c(x))$,
 where $f^\lambda$ is the dimension of $S^\lambda$.
Since the left regular representation of $\Sym_k$ on $\Comp[\Sym_k]$
 is decomposed as
$$
\Comp[\Sym_k]
 \cong \bigoplus_{\lambda \vdash k} \left( S^\lambda \right)^{\oplus f^\lambda},
$$
the trace of $f(L_1) + \cdots + f(L_k)$ on $\Comp[\Sym_k]$ is given by
$$
\sum_{\lambda \vdash k}
 \left( f^\lambda \right)^2 \sum_{x \in \lambda} f(c(x))
 =
(k!)^2 \sum_{\lambda \vdash k}
 \frac{ \sum_{x \in \lambda} f(c(x)) }
      { \prod_{x \in \lambda} h(x)^2 },
$$
because $f^\lambda = k! / \prod_{x \in \lambda} h(x)$.

On the other hand, the trace of the operator $g \in \Sym_k$
 on $\Comp[\Sym_k]$ is equal to $k!$ if $g$ is the identity element
 and $0$ otherwise.
Hence we see that
$$
k! \sum_{\lambda \vdash k}
 \frac{ \sum_{x \in \lambda} f(c(x)) }
      { \prod_{x \in \lambda} h(x)^2 }
$$
is the coefficient of the identity element in $f(L_1) + \cdots + f(L_k)$.
\qed
\end{demo}

Now we recall the Lascoux--Thibon formula, which expresses
 the power-sums of Jucys--Murphy elements as linear combinations 
of the class sums $C_\mu$.
For a partition $\mu$, we denote by $C_\mu$ the sum of
 all permutations with cycle type $\mu$ and put
 $z_\mu = \prod_{i \ge 1} i^{m_i} m_i!$, where $m_i$
 is the multiplicity of $i$ in $\mu$.

\begin{theorem} (Lascoux--Thibon \cite[\S 4]{LT})
Given a partition $\kappa$ of $m$, we define a formal power series
 $\phi_\kappa(t) = \sum_{r \ge 0} \phi_{\kappa,r} t^r/r!$ by
 substituting $q=e^t$ in
$$
\frac{ (1-q^{-1})^{m-1} }
     { m! z_\kappa }
\frac{ \prod_i (q^{\kappa_i} - 1) }
     { q-1 }.
$$
Then we have
$$
L_1^r + \cdots + L_k^r
 =
\sum_{m=1}^{r+1}
\sum_{\substack{\kappa \vdash m \\ \ l(\kappa) \le r-m+2}}
 \phi_{\kappa,r}
 \frac{ z_{\kappa \cup (1^{k-m})} }
      { (k-m)! }
 C_{\kappa \cup (1^{k-m}) }.
$$
\end{theorem}

\begin{corollary}
If $r \ge 1$, then the coefficient of the identity element
 in $L_1^{2r} + \cdots + L_k^{2r}$ is given by
$$
\sum_{p=1}^r \frac{2^p}{((p+1)!)^2} k^{\underline{p+1}}
\sum_{\substack{r_1 + \cdots + r_p = r \\ r_1, \cdots, r_p > 0}}
 \binom{2r}{2r_1, 2r_2, \cdots, 2r_p},
$$
where the inner sum is taken over all $p$-tuples of positive integers
 $(r_1, \cdots, r_p)$ with $r_1 + \cdots + r_p = r$,
 and $\binom{2r}{2r_1, 2r_2, \cdots, 2r_p}$ is the multinomial coefficient.
\end{corollary}

\begin{demo}{Proof}
We consider the coefficient of $C_{(1^k)}$ in the Lascoux--Thibon formula.
If $\kappa = (1)$, then $\phi_{(1)} = 1$ and
 $\phi_{(1),r} = 0$ for $r \ge 1$.
If $\kappa = (1^m)$ with $m \ge 2$, then 
$$
\phi_{(1^m)}(t)
 = \frac{1}{(m!)^2} \left( e^t -2 + e^{-t} \right)^{m-1},
$$
and
$$
\phi_{(1^m),2r}
 =
\frac{2^{m-1}}{(m!)^2} 
\sum_{\substack{r_1 + \cdots + r_{m-1} = r \\ r_1, \cdots, r_{m-1} > 0}}
 \binom{2r}{2r_1, 2r_2, \cdots, 2r_{m-1}},
$$
where the sum is taken over all $(m-1)$-tuples of positive integers
 $(r_1, \cdots, r_{m-1})$ with $r_1 + \cdots + r_{m-1} = r$.
\qed
\end{demo}

Now the proof of (\ref{a2}) is completed by showing the following lemma.

\begin{lemma}
If $r \ge p$, then we have
\begin{equation}
\frac{2^p}{(2p)!}
\sum_{\substack{r_1 + \cdots + r_p = r \\ r_1, \cdots, r_p > 0}}
 \binom{2r}{2r_1, 2r_2, \cdots, 2r_p}
 = h_{r-p}(1^2, 2^2, \cdots, p^2).
\label{a3}
\end{equation}
\end{lemma}

\begin{demo}{Proof}
First we simplify the summation on the left hand side of (\ref{a3}).
We put
\begin{align*}
M_p(r)
 &=
\sum_{\substack{r_1+\cdots+r_p=r \\ r_1, \cdots, r_p \ge 0}}
 \binom{2r}{2r_1, \cdots, 2r_p},
\\
N_p(r)
 &=
\sum_{\substack{r_1+\cdots+r_p=r \\ r_1, \cdots, r_p > 0}}
 \binom{2r}{2r_1, \cdots, 2r_p}.
\end{align*}
(We define $M_0(r) = N_0(r) = 0$.)
It follows from the multinomial theorem that
\begin{align*}
&\sum_{(\ep_1, \cdots, \ep_p) \in \{ 1, -1 \}^p}
 (\ep_1 x_1 + \cdots + \ep x_p)^{2r}
\nonumber\\&\qquad\qquad
=
\sum_{r_1, \cdots, r_p}
 \binom{2r}{r_1, \cdots, r_p}
 \left( x_1^{r_1} + (-x_1)^{r_1} \right)
 \cdots
 \left( x_p^{r_p} + (-x_1)^{r_p} \right),
\end{align*}
where the sum is taken over all $p$-tuples $(r_1, \cdots, r_p)$ of
 non-negative integers with $r_1 + \cdots + r_p=r$.
Substituting $x_1 = \cdots = x_p = 1$, we obtain
$$
M_p(r)
 =
\frac{1}{2^p}
 \sum_{i=0}^p \binom{p}{i} (p-2i)^{2r}.
$$
By applying the Principle of Inclusion--Exclusion, we have
\begin{align*}
N_p(r)
&=
\sum_{j=0}^p (-1)^j\binom{p}{j} M_{p-j}(r)
\\
&=
\frac{1}{2^p}
\left(
 \sum_{k=0}^{\lfloor p/2 \rfloor}
 \left[ \sum_{i=0}^k 2^{2i} \binom{p}{2i} \binom{p-2i}{k-i} \right]
 \left( (p-2k)^{2r} + (-p+2k)^{2r} \right)
\right.
\\
&\!\!\!\!
 -
\left.
 \sum_{k=0}^{\lfloor (p-1)/2 \rfloor}
 \left[ \sum_{i=0}^k 2^{2i+1} \binom{p}{2i+1} \binom{p-2i-1}{k-i}
 \right]
 \left( (p-2k-1)^{2r} + (-p+2k+1)^{2r} \right)
\right).
\end{align*}
By using the Chu--Vandermonde formula (see e.g. \cite[Cor.~2.2.3]{AAR}),
 we see that
\begin{align*}
&\text{if $0 \le 2k \le 2p$,}\\
&\qquad\qquad\sum_{i=0}^k 2^{2i} \binom{p}{2i} \binom{p-2i}{k-i}
 =\frac{ p^{\underline{k}} }{ k! }
 \sum_{i=0}^k
   \frac{ (p-k)^{\underline{i}} k^{\underline{i}} }
        { i! (i-1/2)^{\underline{i}} }
 =\binom{2p}{2k},
\\
&\text{if $0 \le 2k+1 \le 2p$,}\\
&\qquad\qquad\sum_{i=0}^k 2^{2i+1} \binom{p}{2i+1} \binom{p-2i-1}{k-i}
 =\frac{ 2 p^{\underline{k+1}} }{ k! }
\sum_{i=0}^k
 \frac{ (p-k-1)^{\underline{i}} k^{\underline{i}} }
      { i! (i+1/2)^{\underline{i}} }
 =\binom{2p}{2k+1}.
\end{align*}
Therefore we conclude that
$$
N_p(r)
 =
\frac{2}{2^p}
 \sum_{i=1}^p (-1)^{p-i} \binom{2p}{p-i} i^{2r}.
$$

Now we are in position to complete the proof of (\ref{a3}) by
 using generating functions.
The generating function of the right hand sides is
\begin{align*}
\sum_{r=p}^\infty \frac{2^p}{(2p)!} N_p(r) z^{r-p}
 &=
\sum_{r=p}^\infty
 \frac{2}{(2p)!}
 \left( \sum_{i=1}^p (-1)^{p-i} \binom{2p}{p-i} i^{2r} \right)
 z^{r-p}
\\
 &=
 \sum_{i=1}^p (-1)^{p-i} \frac{ 2 i^{2p} }{ (p-i)!(p+i)! }
 \frac{1}{1 - i^2 z}.
\end{align*}
By considering the partial fraction expansion, we see that
$$
\sum_{i=1}^p (-1)^{p-i} \frac{ 2 i^{2p} }{ (p-i)!(p+i)! }
 \frac{1}{1 - i^2 z}
 =
\prod_{i=1}^p \frac{1}{1-i^2 z},
$$
which is the generating function
 $\sum_{r=p}^\infty h_{r-p}(1^2, \cdots, p^2) z^{r-p}$.
This completes the proof of (\ref{a3}).
\qed
\end{demo}



\begin{thebibliography}{99}

\bibitem{Nek} 
N. Nekrasov, Seiberg-Witten Prepotential
from Instanton Counting,
Adv. Theor. Math. Phys. 7 (2004) 831,
{\tt arXiv:hep-th/0206161}.

\bibitem{FP}
R. Flume and R. Poghossian,
An algorithm for the microscopic evaluation of the coefficients of the
Seiberg-Witten prepotential,
Int. J. Mod. Phys. {\bf A 18} (2003) 2541,
{\tt arXiv:hep-th/0208176}.

\bibitem{BFMT}
U. Bruzzo, F. Fucito, J.M. Morales and A. Tanzini,
Multi-instanton calculus and equivariant cohomology,
JHEP {\bf 05} (2003) 054,
{\tt arXiv:hep-th/0211108}.

\bibitem{LMN}
A. Losev, A. Marshakov and N. Nekrasov,
Small Instantons, Little Strings and Free Fermions,
{\tt arXiv:hep-th/0302191}.

\bibitem{NO}
N. Nekrasov, A. Okounkov, 
Seiberg-Witten Prepotential and Random Partitions,  \hfill\break
{\it The unity of mathematics,} Progr. Math. {\bf 244} (2006) 525-596, \hfill\break
{\tt arXiv:hep-th/0306238}.

\bibitem{NY1}
H. Nakajima and K. Yoshioka,
Instanton Counting on Blowup I,
Invent. Math {\bf 162 no. 2} (2005) 313-355,
{\tt arXiv:math.AG/0306198}.

\bibitem{FFMP}
R. Flume, F. Fucito, J.M. Morales and R. Poghossian,
Matone's Relation in the Presence of Gravitational Couplings,
JHEP {\bf 04} (2004) 008,
{\tt arXiv:hep-th/0403057}.


\bibitem{FMPT}
F. Fucito, J.M. Morales, R. Poghossian and A. Tanzini,
${\cal N}=1$ superpotentials from Multi-Instanton Calculus,
JHEP {\bf 01} (2006) 031,
{\tt arXiv:hep-th/0510173}.

\bibitem{IK1} A. Iqbal and A.-K. Kashani-Poor, 
Instanton Counting and Chern-Simons Theory,
Adv. Theor. Math. Phys. {\bf 7} (2004) 457,
{\tt arXiv:hep-th/0212279}.

\bibitem{IK2} A. Iqbal and A.-K. Kashani-Poor, 
$SU(N)$ Geometries and Topological String Amplitudes,
Adv. Theor. Math. Phys. {\bf 10 no. 1} (2006) 1-32,
{\tt arXiv:hep-th/0306032}.

\bibitem{EK1} T. Eguchi and H. Kanno, 
Topological Strings and Nekrasov's Formulas,
JHEP  {\bf 12} (2003) 006,
 {\tt arXiv:hep-th/0310235}.

\bibitem{EK2} T. Eguchi and H. Kanno, 
Geometric transitions, Chern-Simons gauge theory  
and Veneziano type amplitudes,
 Phys. Lett.  {\bf B 585} (2004) 163-172, \hfill\break
 {\tt arXiv:hep-th/0312234}.

\bibitem{Zhou} Jian Zhou, 
Curve Counting and Instanton Counting,
{\tt arXiv:math.AG/0311237}.

\bibitem{GV}
R. Gopakumar and C. Vafa,
On the Gauge Theory/Geometry Correspondence,
Adv. Theor. Math. Phys. {\bf 3} (1999) 1415,
{\tt arXiv:hep-th/9811131}.

\bibitem{KKV}
S. Katz, A. Klemm and C. Vafa,
Geometric Engineering of Quantum Field Theories,
Nucl. Phys. {\bf B 497} (1997) 173,
 {\tt arXiv:hep-th/9609239}.

\bibitem{OP1}
A. Okounkov and R. Pandharipande,
Gromov-Witten Theory, Hurwitz Theory and Completed Cycles,
Ann. of Math. {\bf 163 no. 2} (2006) 517-560,  \hfill\break
{\tt arXiv:math.AG/0204305}.

\bibitem{OP2}
A. Okounkov and R. Pandharipande,
The Equivariant Gromov-Witten Theory of ${\bf P}^1$,
Ann. of Math. {\bf 163 no. 2}  (2006) 561-605,
{\tt arXiv:math.AG/0207233}.

\bibitem{KMT} A.~Klemm, M.Mari\~no and S.~Theisen,
Gravitational Corrections in Supersymmetric Gauge
Theory and Matrix Models, JHEP {\bf 03} (2003) 051,
{\tt arXiv:hep-th/0211216}.

\bibitem{DST} R.~Dijkgraaf, A.~Sinkovics and M.~Tem\"urhan,
 Matrix Models and Gravitational Corrections,
Adv.Theor.Math.Phys. {\bf 7} (2004) 1155-1176,
{\tt arXiv:hep-th/0211241}.


\bibitem{BFFL}
M. Bill\'o, M. Frau, F. Fucito and A. Lerda,
Instanton Calculus in $R-R$ Background and the Topological String,
JHEP {\bf 11} (2006) 012,
{\tt arXiv:hep-th/0606013}.


\bibitem{MSS}
G.W. Moore, N. Seiberg and M. Staudacher,
From loops to states in 2-D quantum gravity
Nucl.\ Phys.\  B {\bf 362} (1991) 665.

\bibitem{ESZ}
J.K. Erickson, G.W. Semenoff and K. Zarembo,
Wilson loops in N = 4 supersymmetric Yang-Mills theory,
Nucl.\ Phys.\  B {\bf 582} (2000) 155,
{\tt arXiv:hep-th/0003055}.

\bibitem{DG}
N. Drukker and D.J. Gross,
An exact prediction of N = 4 SUSYM theory for string theory,
J.\ Math.\ Phys.\  {\bf 42} (2001) 2896,
{\tt arXiv:hep-th/0010274}.

\bibitem{BH}
G. Bertoldi and T.J. Hollowood,
Large $N$ gauge theories and topological cigars,
{\tt arXiv:hep-th/0611016}.

\bibitem{MN}
A. Marshakov and N. Nekrasov,
Extended Seiberg-Witten Theory and Integrable Hierarchy,
{\tt arXiv:hep-th/0612019}.

\bibitem{BS}
L. Baulieu and I.M. Singer, Topological Yang-Mills Symmetry, 
Nucl. Phys. Proc. Suppl. {\bf 5B} (1988) 12-19.

\bibitem{Kan}
H. Kanno,  Weil Algebra Structure and Geometrical Meaning 
of BRST Transformation in Topological Quantum Field Theory,
Z. Phys.  {\bf C43} (1989) 477-484.

\bibitem{BGV}
N. Berline, E. Getzler and M. Vergne,
{\it Heat Kernels and Dirac Operators}, 
Springer-Verlag, (1996).

\bibitem{GS}
V.W. Guillemin and S. Sternberg,
{\it Supersymmetry and Equivariant de Rham Theory},
Springer-Verlag, (1999).

\bibitem{Nak}
H. Nakajima, 
{\it Lectures on {H}ilbert schemes of points on surfaces},
{University Lecture Series}, {\bf 18}, 
American Mathematical Society, (1999).

\bibitem{CDSW}
F. Cachazo, M.R. Douglas, N. Seiberg and E. Witten,
Chiral Rings and Anomalies in Supersymmetric Gauge Theory,
JHEP {\bf 12} (2002) 071,
{\tt arXiv:hep-th/0211170}.

\bibitem{NY2}
H. Nakajima and K. Yoshioka,
Lectures on instanton counting,
In:  {\it Algebraic structures and moduli spaces}, CRM Proc. Lecture Notes, {\bf 38},
31-101, {Amer. Math. Soc.}, (2004).

\bibitem{LQW}
Wei-Ping Li, Zhanbo Qin and Weiqiang Wang,
Hilbert Schemes, Integrable Hierarchies and Gromov-Witten Theory,
Internat. Math. Res. Notices {\bf 40} (2004) 2085-2104,
{\tt arXiv:math.AG/0302211}.

\bibitem{Mac}
I. G. MacDonald,
{\it Symmetric Functions and Hall Polynomials}, (Second Edition),
Oxford University Press, (1995), [Chap. 1.4, Example 3].


\bibitem{AAR}
G.~E.~Andrews, R.~Askey, and R.~Roy, 
{\it Special Functions},
Cambridge Univ. Press, (1999).

\bibitem{J}
A.-A.~A.~Jucys,
Symmetric polynomials and the center of the symmetric group ring,
Rep. Mathematical Phys. {\bf 5} (1974), 107--112.

\bibitem{LT}
A.~Lascoux and J.-Y.~Thibon, 
Vertex operators and the class algebras of symmetric groups, 
J. Math. Sci. (N. Y.) {\bf 121} (2004), 2380--2392, 
{\tt arXiv:math.CO/0102041}.

\bibitem{Mur}
G.~E.~Murphy,
A new construction of Young's seminormal representation of the symmetric groups,
J. Algebra {\bf 69} (1981), 287--297.

\end{thebibliography}
\end{document}